\documentclass[preprint,showpacs,preprintnumbers,amsmath,amssymb]{revtex4}

\usepackage{etex}
\usepackage{graphicx}
\usepackage{dcolumn}
\usepackage{bm}
\usepackage{amssymb,amsthm,amscd,amsbsy,array}
\usepackage{amsmath,dsfont}
\usepackage{soul} 
\usepackage{graphics,xcolor}

\numberwithin{equation}{section}

\newcommand{\half}{{\scriptstyle{\frac{1}{2}}}}
\def\2{{\half}}
\newcommand{\const}{\mathop{\rm const}\nolimits}

\def\p{{\partial}}

\def\bp{{\bm{p}}}

\def\tbx{\tilde{\bm{x}}}

\def\bnabla{\mbox{\boldmath$\nabla$}}
\def\br{{\bm{r}}}

\def\bg{{\bm{g}}}

\def\bE{{\bm{E}}}
\def\bB{{\bm{B}}}
\def\bb{{\bm{b}}}

\def\bnabla{{\bm{\nabla}}}

\def\bp{{\bm{p}}}

\def\bs{{\bm{s}}}

\def\hbp{{\widehat{\bm{p}}}}

\def\bj{{\bm{j}}}
\def\bx{{\bm{x}}}

\def\beq{\begin{equation}}
\def\eeq{\end{equation}}
\def\beqa{\begin{eqnarray}}
\def\eeqa{\end{eqnarray}}
\def\nn{\nonumber}
\def\barray{\left(\begin{array}}
\def\earray{\end{array}\right)}
\def\barraynb{\begin{array}}
\def\earraynb{\end{array}}

\def\smallover#1/#2{\hbox{$\textstyle\frac{#1}{#2}$}} %

\newcommand{\cE}{{\mathcal{E}}}

\newcommand{\belle}{\boldsymbol{\ell}}

\usepackage{color}

\begin{document}

\preprint{ arXiv~:~1506.05008v4 [hep-th]}

\title{Anomalous properties of spin-extended chiral fermions
}

\author{
M. Elbistan$^{1}$\footnote{mailto:elbistan@itu.edu.tr},
P. A. Horvathy$^{1,2}$\footnote{mailto:horvathy@lmpt.univ-tours.fr},
}

\affiliation{
${}^1$
Institute of Modern Physics, Chinese Academy of Sciences, Lanzhou, (China) 
\\
$^2$Laboratoire de Math\'ematiques et de Physique
Th\'eorique,
Universit\'e de Tours,  
(France)
}

\date{\today}

\begin{abstract}The spin-extended semiclassical chiral fermion (we call the  S-model), which had been used to derive the twisted Lorentz symmetry of the  ``spin-enslaved" chiral fermion (we call the c-model)  is equivalent to the latter in the free case, however coupling to an external electromagnetic field yields nonequivalent systems. The difference is highlighted by the inconsistency of  spin enslavement within the spin-extended framework. The  S-model exhibits nevertheless similar though slightly different anomalous properties as the usual c-model does.
The natural Poincar\'e symmetry of the free model remains unbroken if the Pfaffian invariant vanishes, i.e., when the electric and magnetic fields are orthogonal, $\bE\cdot\bB=0$ as in the Hall effect. \\ 
\end{abstract}


\maketitle


\section{Introduction}

The semiclassical chiral model (we call here the \emph{c-model})
allows for a derivation of the chiral magnetic effect and the chiral anomaly,
respectively, bypassing complicated quantum calculations
\cite{SonYama2, StephanovYin,Stone,QunWang,Manuel}. 
The free c-model,  which has no genuine spin degree of freedom,
carries  a curious ``twisted'' Lorentz symmetry \cite{ChenSon,DHchiral,DEHZ,deAzcarraga,Kosinski}, conveniently derived by relating it to  Souriau's massless spinning particle  
\cite{SSD}. The latter (we call the \emph{S-model}), carries a mass-zero, spin-$s$ Poincar\'e symmetry. 
Compared to the c-model, the S-model has two additional degrees of freedom represented by an ``unchained'' spin vector, $\bs$, whose projection onto the momentum is fixed, 
$
\bs\cdot\hbp=s
$ \cite{DHchiral,DEHZ}. 
Free spin can however be ``enslaved''  to the momentum, 
\beq 
\bs=s\,\hat{\bp},
\label{enslavement}
\eeq
 by a suitable  ``Wigner-Souriau translation'',
which embeds the free c-model into that of Souriau \cite{SSD},
making them equivalent \cite{DHchiral,DEHZ}. 

The c- and S-models are no longer equivalent, though,
 when the systems are put into a field, as highlighted by the explicit solution
presented in Sec. 5 A of \cite{DHchiral}. In particular, \emph{spin can no longer be consistently enslaved} within the S-model
\cite{DHchiral}.

In this Letter we show that the minimally coupled S-model, although nonequivalent to the c-model, admits nevertheless similar transport properties, see eqns (\ref{SLiouville})-(\ref{Scurrent})-(\ref{Sanomaly}) below. 

We formulate our results within the framework of Souriau \cite{SSD,DHchiral}. To make our  paper more self contained, we remind the reader of some basic facts while referring  to these references for details.

The ultimate description of a mechanical system is provided by its \emph{space of motions}, $M$, which is a \emph{symplectic} [and therefore an \emph{even-dimensional}] manifold; its symplectic form, $\omega$, is regular. A symmetry group acts on $M$ by preserving its symplectic form, $\omega$. If the symplectic action is transitive, then $(M,\omega)$ is identified with a \emph{coadjoint orbit carrying its canonical symplectic form}. 

Conversely, one can start by constructing such an orbit and then seek a physical interpretation for it. For this end, it is convenient to use what Souriau calls an \emph{evolution space}, $V$, which is endowed with a closed  two-form $\sigma$ of constant rank $r\geq1$. The distribution provided by $\ker\sigma$ is integrable, and its characteristic leaves are identified with the classical motions of the system. The space of motions is the quotient of the evolution space by the characteristic foliation of $\sigma$, 
\beq
(M,\omega)=(V,\sigma)/\ker\sigma.
\label{SoM}
\eeq
Thus dim $V$ = dim $M + r$. The points of $M$ are be labeled by \emph{constants of the motions}.

The concrete realization of this abstract framework in our context is summarized in section  \ref{mincoup} below.

\goodbreak

\section{The c-model}

We first summarize some aspects of the  c-model we will need to 
be compared with those in the S-model.
In Souriau's framework \cite{SSD,DHchiral,DEHZ} the  model 
 can be described by a $7-$dimensional evolution space $V^7$ with coordinates $(\bx,\bp,t)$, endowed with the closed $2-$form $\sigma_c=\omega_c-dh\wedge dt$ with one-dimensional kernel,
where  the symplectic form and the Hamiltonian are,
\beq
\label{comega}
\omega_c=dp_i\wedge dx_i+\frac{e}{2}\epsilon_{ijk}B_idx_j\wedge dx_k
-\frac{1}{4|\bp|^3}\epsilon_{ijk}p_idp_j\wedge dp_k,
\qquad
h=|\bp|+e\phi,
\eeq
respectively, where $-\bnabla\phi=\bE$ \cite{SSD,DHchiral}.
The Poisson brackets are therefore 
\begin{eqnarray}
\{x_i,x_j\}=\varepsilon_{ijk}\displaystyle\frac{b_k}{1+e\bb\cdot\bB},
\quad
\{x_i,p_j\}=\displaystyle\frac{\delta_{i j}+eB_ib_j}{1+e\bb\cdot\bB},
\quad
\{p_i,p_j\}=-\varepsilon_{ijk}\displaystyle\frac{eB_k}{1+e\bb\cdot\bB}\,,
\quad
\label{cPB}
\end{eqnarray}
where $\bb=s\,{\hbp}/|\bp|^2, \, \hbp=\bp/|\bp|$ is the ``Berry monopole'' of strength $s$ in the momentum space \cite{DHHMS}. Usually  $s=1/2$ \cite{SonYama2, StephanovYin,Stone,QunWang,Manuel}. 
In the denominator  we recognize here the square-root of the determinant of the symplectic matrix $\omega_c$,
$1+e\bm{b}\cdot\bB=\sqrt{\det(\omega_c)}=D_c$. The system is  regular 
when $D_c\neq0$. 
The Hamilton equations 
\begin{subequations}
\label{ceqs}
\begin{align}
D_c\,\frac{d\bx}{dt}&=\hat{\bp}+e\bm{E}\times\bm{b}+e(\bm{b}\cdot\hat{\bp})\bB,
\\[6pt]
D_c\,\frac{d\bp}{dt}&=e\bm{E}+e\hat{\bp}\times\bB+e^2(\bm{E}\cdot\bB)\bm{b},
\end{align}
\end{subequations}
reproduce eqns. (14)-(15) in \cite{StephanovYin} and correspond to
the one-dimensional kernel of $\sigma_c$ \cite{SSD,DHchiral,DHHMS}. Factoring out the kernel
yields the $6$ dimensional space of motions (identified here with the phase space) \cite{SSD,DHchiral}. Particular solutions putting in evidence the role of the anomalous velocity in (\ref{ceqs}) were studied in
\cite{ZhH}.

The invariant volume element is the $6/2=3$rd power of the symplectic form $\omega_c$ \cite{SSD,DHHMS} and its pull-back to the evolution space $V^7$ is the 3rd power of $\sigma_c$,
\beq
d{\rm V}_c=D_c\,d^3\bp d^3\bx.
\label{volel}
\eeq
Then Liouville's theorem takes the anomalous form \footnote{
Here and elsewhere, the use of the Dirac delta which ``lives in the hole'' $\bp=0$ is somewhat abusive. It could be bypassed however
by excising a small sphere in momentum space around its origin $\bp=0$ and then letting the radius go to zero.}
\beq
\frac{\p{D_c}}{\p t}+\frac{\p({D_c}\dot{\bx})}{\p \bx}
+\frac{\p({D_c}\,\dot{\bp})}{\p \bp}= (\bE\cdot\bB)\,\bnabla_{\bp}\cdot\left(\frac{\hbp}{2|\bp|^2}\right)
=
2\pi e^2(\bE\cdot\bB)\, \delta^3(\bp).
\label{cLiouville}
\eeq

Let $f(\bx,\bp,t)$ be a distribution on the phase space which we assume to satisfy the collision-less Boltzmann equation
$
\p_tf+\p_{\bx}f\,\dot{\bx}
+\p_{\bp}f\,\dot{\bp}=0.
$
The current density is,  
\beqa
\bj =\displaystyle\int{f\,\dot{\bx}}\, D_c\frac{d^3\bp}{(2\pi)^3}
=
\int{f}\hbp\frac{d^3\bp}{(2\pi)^3}+
e\bB\int\frac{f}{2|\bp|^2}\frac{d^3\bp}{(2\pi)^3}\,
&+&
e\bE\times\int\frac{f\hbp}{2|\bp|^2}\frac{d^3\bp}{(2\pi)^3}.
\label{ccurrent}
\eeqa
The first term on the r.h.s. is the normal current, the second one represents the \emph{chiral magnetic effect} (CME) and the last 
one is the anomalous Hall current \cite{StephanovYin}.
Defining the particle density as
$ 
\rho(\bx,t)=\displaystyle\int\!fD_c\frac{d^3\bp}{(2\pi)^3}
$
yields the anomalous continuity equation (referred to as the \emph{chiral anomaly}),
\beq
\p_t\rho+\bnabla\cdot\bj=\frac{e^2}{4\pi^2}(\bE\cdot\bB)f_0,
\label{canomaly}
\eeq 
where $f_0$ is the value of the distribution function at $\bp=0$
\cite{StephanovYin,SonYama2,Stone,QunWang}.

\section{The  massless spinning model, minimally coupled to an e.m. field }\label{mincoup}

The  evolution space of the S-model, $V^9$, has, w.r.t. to the c-model, two additional degrees of freedom represented by the spin vector, $\bs$,
whose projection onto the momentum is fixed, 
$
\bs\cdot\hbp=s
$ 
\cite{DHchiral}. 
However, the kernel of the free two-form $\sigma_c$ which yields the equations of motion is now \emph{$3$-dimensional} spanned by Wigner-Souriau translations, yielding, once again, a $6$-dimensional  space of free motions \cite{SSD,DHchiral}.

\begin{figure}[ht]\vskip-2mm
\begin{center}
\includegraphics[scale=.67]{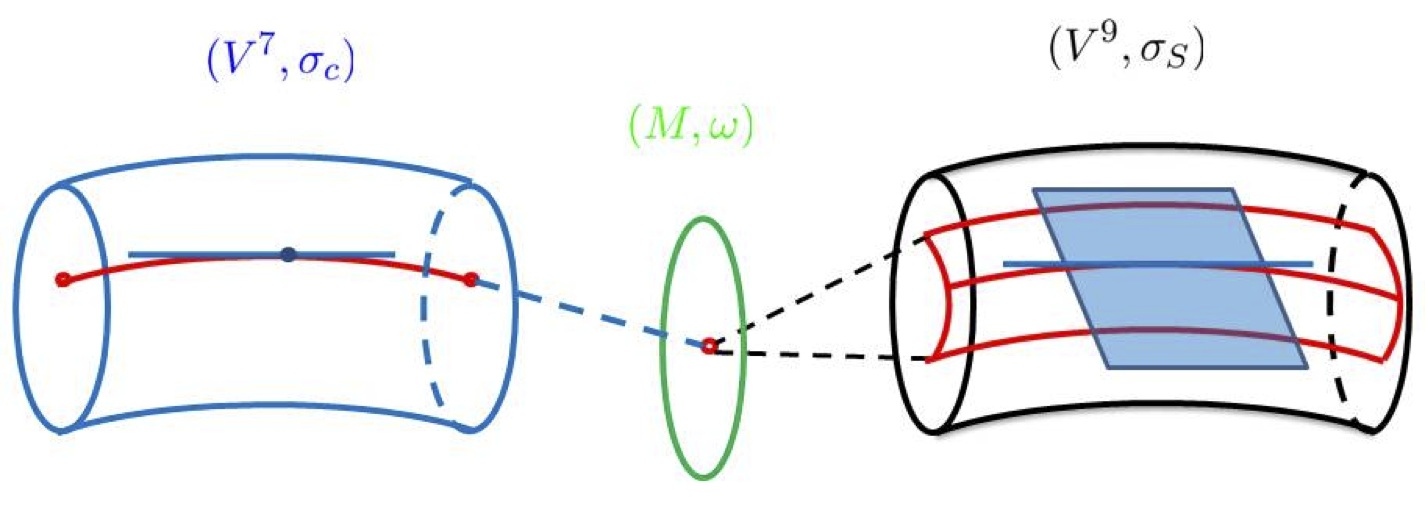}\;
\end{center}\vskip-10mm
\caption{ {\it The free motions of the c-model are described by curves in the $7$-dimensional evolution space, whereas the  motions of the S-model are 3-dimensional surfaces lying in $9$-dimensional evolution space $V^9$. Factoring out the motions yields, however, the \emph{same} space of motions, $(M,\omega)$, for both systems.}
} 
\label{figchirmassless}
\end{figure}

It is worth, at this point, to compare the c- and S-models. In the c-model the evolution space, $V^7$, is $7$-dimensional and the motions are curves, which are tangent to the $1$-dimensional kernel of $\sigma_c$. In the S-model instead, the evolution space, $V^9$, is $9$-dimensional; the kernel is $\sigma_S$ is 3 dimensional. The respective spaces of free motion are, therefore, $6$ dimensional in both cases. The key point is that factoring out the respective motions yields, in both cases, the \emph{same} space of motions, $(M,\omega)$, as illustrated on Fig. \ref{figchirmassless}.

 Coupling to an external electromagnetic field is introduced through Souriau's minimal coupling schema \cite{SSD}, which requires to add to the free form $\sigma$ [$e$-times] the electromagnetic field tensor, written in terms of the ``true'' position, $\br$, -- the one which transforms in the usual way under a Lorentz boost  \cite{DHchiral}. 

The S-model is thus described by a $9$-dimensional evolution space $V^9$
with coordinates $\br,\bp\neq0,\bs, (\,\bs\cdot\hbp=s=\half)$, endowed with the  closed $2$-form $\sigma_S$ \cite{DHchiral}. The Hamiltonian is still of the form $h=|\bp|+e\phi$,  
remember however that \emph{the potential (assumed static) is now a function of the ``true" position}
$\br$, $\phi=\phi(\br)$. Spelling out in $3+1$ dimensional notations the constraints which define the spin-extended evolution space and the symplectic form, 
\begin{subequations}
\begin{align}
P_\mu{}P^\mu=0,\quad
S_{\mu\nu}P^\nu=0,\quad
\half{S}_{\mu\nu}{S}^{\mu\nu}=s^2,
\label{3.4}
\\
\sigma=
-dP_\mu\wedge{}dR^\mu
-\frac{1}{2s^2}\,d{S}^\mu_{\;\lambda}\wedge{S}^\lambda_{\;\rho}\,d{S}^\rho_{\;\mu}
+\frac{e}2 F_{\mu\nu}\,
dR^\mu\wedge dR^\nu,
\label{5.3}
\end{align}
\end{subequations}
cf. eqns. \# (5.3) and \# (3.4) in  \cite{DHchiral},
a long but straightforward calculation using also the constraints in eqn. (\ref{3.4}) yields the  complicated-looking symplectic form $\sigma_S=\omega_S-dh\wedge dt$,
\begin{eqnarray}
\label{sigmag03+1}
\omega_S
&=&dp_i\wedge dr_i+\frac{e}{2}\epsilon_{ijk}B_idr_j\wedge dr_k
\nonumber\\
&+&\frac{2\bs^2}{|\bp|^2}\big[\epsilon_{ijk}s_i+2\hat{p}_j(\hat{\bp}\times\bs)_k\big]dp_j\wedge dp_k
\nonumber \\
 &+&
2\big[\epsilon_{ijk}(\frac{\hat{p}_i}{2}-s_i)+2(\hat{\bp}\times\bs)_j\hat{p}_k \big]ds_j\wedge ds_k\quad
\label{Somega}
\\
&-&\frac{2}{|\bp|}\big[ \epsilon_{ijk}(\bs^2\hat{p}_i+\frac{1}{2}s_i)+(\hat{\bp}\times\bs)_js_k+(\hat{p}_j-s_j)(\hat{\bp}\times\bs)_k \big]dp_j\wedge ds_k.
\nonumber
\end{eqnarray} 
If spin could be enslaved, $\bs = s\hbp$ as in the free case, terms with $\hat{\bp}\times\bs$ would drop out. As we show below however, \emph{spin enslavement is consistent with the S-dynamics  only in the free case} in that (\ref{reducedomega})  would reduce to (\ref{comega})  in the free, but \emph{not} in the coupled case, when the two models are radically different.

The  rather weirdly-looking equations of motion calculated from the kernel of $\sigma_S$,
\begin{subequations}
\begin{align}
(\hbp\cdot\bB)\frac{d\br}{dt}&=\bB-\hat{\bp}\times\bE,
\label{Seqsa}
\\
(\hbp\cdot\bB)\frac{d\bp}{dt}&=e(\bE\cdot\bB)\hat{\bp},
\label{Seqsb}
\\
(\hbp\cdot\bB)\frac{d\bs}{dt}&=\bp\times\bB-\bp\times(\hat{\bp}\times \bE),
\label{Seqsc}
\end{align}
\end{subequations}  
are similar to but different from the analogous equations, (\ref{ceqs}), in the c-model. Note in particular the absence of the
usual momentum on the rhs. of the velocity relation which is, so to say, 
``purely anomalous''. The upper two equations here are decoupled from the
lowest one, so that the 
space-time motion  does not depend on the spin at all. We record for later use that the \emph{direction} of the momentum is a constant of the motion, $d\hbp/dt=0$. Eqn  (\ref{Seqsc}) shows, moreover, that
$\dot{\bs}=\bp\times\dot{\br}$. Thus, although spin is not more enslaved,
its motion is entirely determined by that in space-time.

The eqns (\ref{Seqsa}-c) are valid under the regularity assumptions  \cite{DHchiral}
\beq
\label{assumptions}
(i)\;\;\hat{\bp}\cdot \bB\neq 0,
 \qquad 
(ii) \;\; \bs\cdot(\bB-\hat{\bp}\times \bE)\neq 0.
\eeq
These conditions are preserved by the dynamics, as shown by using  
 the equations of motion. The first of these conditions will be interpreted below as the non-vanishing of the system's determinant.
 
It is worth stressing that the neither the free equations of motion nor the free Souriau form $\sigma_S^{free}$  can be recovered by simply letting the fields go to zero. This limit is in fact a singular one as seen from eqn. (\ref{assumptions}) and is highlighted by the fact that turning off the fields converts the one-dimensional motion curves into 3-dimensional surfaces, cf. \cite{DHchiral}.

As proved in \cite{DHchiral} and seen also directly, the helicity condition $\bs\cdot\hbp=\half$ is  consistent with the coupled equations of motion.

\section{``Dynamical'' Poincar\'e symmetry}

Before turning to study the transport properties, we would like to point out a rather curious fact. For the {free} S-model, the  angular momentum vector, 
\beq
\belle=
 \br\times\bp\,+\bs,
 \label{Sangmom}
\eeq 
is plainly conserved \cite{DHchiral}. Now  for arbitrary constant electric and magnetic fields,
 $\bE$ and $\bB$, the eqns of motion  imply that various terms cancel, leaving us with
\beq
\label{ellem}
\frac{d\belle}{dt}=e\frac{(\bE\cdot\bB)}{\hat{\bp}\cdot \bB}\,\br\times\hat{\bp},
\eeq
[assuming the determinant does not vanish, $\hat{\bp}\cdot \bB\neq 0$]. Therefore (\ref{Sangmom}) is not conserved in general, as expected. A  surprising observation is,  though, that 
when  $\bE$ and $\bB$ are orthogonal so that \emph{the Pfaffian invariant  vanishes}, 
$-\smallover1/4({\star}F.F)=-\bE\cdot\bB=0$,
then 
\emph{all three components of the angular momentum are
 conserved} and 
the system has a full rotational symmetry. %

For comparison, the conserved angular momentum in the $c$-model is
$ 
\belle_c=
 \br\times\bp\,+\half\,\hbp,
$  
i.e., the spin contribution is enslaved to the momentum. Our  calculation leading to (\ref{Sangmom}) shows, however, that $\belle_c$ is \emph{not} conserved in the minimally coupled S-model, $\dot{\belle}_c\neq0$, 
and it is  the ``unchained component'' $\bs-\half\,\hbp$ of spin which is restores angular momentum conservation. 
Further aspects of the angular momentum for  are reviewed in \cite{BliokhRev}.

The  unbroken rotational symmetry in Hall-type crossed e.m. fields we found here above can actually be extended into a full Poincar\'e symmetry \footnote{We are grateful to Christian Duval for informing us of this  \cite{CDPriv}. Below we reproduce his proof with his kind permission.}. 
The equations of motion
 (\ref{Seqsa}-c) imply, in $4$D notations, that the quantity
\beq
\label{Pi}
\Pi^\mu = P^\mu + e F^{\mu}_{\;\nu}R^\nu
\eeq
reminiscent of  ``magnetic translations'' in the massive Landau problem is conserved, and
 a  short calculation  shows that 
\beq
\dot{P}^\mu=e\frac{({\star}F.F)}{2S.F}\,W^{\mu}
\quad\hbox{and}\quad
\dot{M}^{\mu\nu}=e\frac{({\star}F.F)}{2S.F}\,\big(R^{\mu}W^{\nu}-R^{\nu}W^{\mu}\big),
\label{PdotMdot}
\eeq
where 
$W_\sigma=\half\,\epsilon_{\sigma\mu\nu\rho}M^{\mu\nu}P^{\rho}$ is the generalized Pauli-Lubanski vector \cite{EHZD}. 
Therefore  when the Pfaffian vanishes, both the linear and the Lorentz momenta are constants of the motion, 
\beq
P=\const \qquad\hbox{and}\qquad M=\const
\qquad
\hbox{if}
\qquad
{\star}F.F=0
\label{M=const}
\eeq
i.e., the \emph{full Poincar\'e momentum $(M,P)$ is conserved} extending what we had found for the angular momentum.

\section{Transport properties}
 
Turning to the transport properties, we 
choose $\bB\neq0$ to point into the 3rd direction, $\bB=B\,\hat{\bf z}$ and eliminate
one component of the spin vector $\bs$, say $s_3=\frac{1}{\hat{p}_3}\big(\frac{1}{2}-s_1\hat{p}_1-s_2\hat{p}_2\big)$, leaving us with  the  $8$ independent coordinates $\br, \bp, s_1, s_2$. 
 Then a tedious calculation yields
\beqa
\barraynb{lll}
\omega_S&=&dp_i\wedge dr_i+\frac{e}{2}\epsilon_{ijk}B_idr_j\wedge dr_k
+\displaystyle\frac{s_3}{2|\bp|p_3}\epsilon_{ijk}\hat{p}_idp_j\wedge dp_k
\\[6pt]
&-&
\displaystyle\frac{1}{p_3}\big[\hat{p}_1\hat{p}_2dp_1\wedge ds_1+(1-\hat{p}_1^2)dp_1\wedge ds_2-(1-\hat{p}_2^2)dp_2\wedge ds_1\big]
\\[6pt]
&-&
\displaystyle\frac{1}{p_3}\big[-\hat{p}_1\hat{p}_2dp_2\wedge ds_2 +\hat{p}_2\hat{p}_3dp_3\wedge ds_1-\hat{p}_1\hat{p}_3dp_3\wedge ds_2\big]\,.
\earraynb
\label{reducedomega}
\end{eqnarray}
Note that no $ds_j\wedge ds_k$ terms show up in (\ref{reducedomega}). 
Then a lengthy calculation yields the determinant of the symplectic form,
\beq
\label{Sdet}
D_S\equiv\sqrt{\det (\omega_S)}= \frac{e\,\hat{\bp}\cdot\bB}{|\bp|^2\,\hat{p}_3}=
\frac{eB}{|\bp|^2}\,.
\eeq
 This result is surprisingly simple and somewhat unexpected in that it
 does \emph{not} involve the spin.
Note  that, compared to (\ref{volel}), (\ref{Sdet}) has a ``naked'' $\bb\cdot\bB$ term, but the normal ``$1$'' is missing. 

When $D_S\neq0$ the system is regular; the motions follow curves (world lines) so that the space of motions is  $8$-dimensional with coordinates
$(\br,\bp,s_1,s_2)$.  

What happens when $\hbp\cdot\bB\to0$ ? The determinant goes to zero, $D_S\to0$, and the
 system becomes singular, necessitating reduction,  analogous to what happens for an ``exotic particle''
 in the plane \cite{Exotic}. The characteristic world lines degenerate into three dimensional world sheets~: the space of motions of the free massless spinning particle is $6$ dimensional, with coordinates
$(\br,\bp)$ alone \cite{SSD,DHchiral,DEHZ}.

In the regular case $D_S\neq0$ we assume henceforth, the Liouville theorem takes now the form ($a=1,2$) [15],
\beqa
\frac{\p{D_S}}{\p t}+
\frac{\p({D_S}\,\dot{\br})}{\p \br}
+\frac{\p({D_S}\,\dot{\bp})}{\p \bp}
+\frac{\p({D_S}\,\dot{s_a})}{\p s_a}
&=&
e^2(\bE\cdot\bB)\,{\bnabla}_\bp\cdot\big(\frac{\hat{\bp}}{|\bp|^2\hat{p}_3}  \big)
=
e^2\frac{(\bE\cdot\bB)}{\hat{p}_3}\,4\pi\delta^3(\bp). 
\qquad \quad
\label{SLiouville}
\eeqa

This result only differs from (\ref{cLiouville}) in a factor
$2$ and in the factor $\hat{p}_3^{-1}$ which is in fact a constant of the motion, as we noted earlier. 
Paradoxically, the right hand side here, in this \emph{spin-extend model, is
independent of the spin}, as long as the latter does not vanish -- whereas
 the ``spin-enslaved'' c-model 
(\ref{cLiouville})  has a spin-remnant, namely the factor $s=1/2$.
\footnote{In the Hall-type setup studied in \cite{DHchiral} the Pfaffian vanishes, $\bE\cdot\bB=0$, and no anomaly arises.}.

It can be inferred from the equations of motion that  the helicity constraint  $\hbp\cdot\bs=s$ is preserved by the S-dynamics. However the 
explicit solutions found in \cite{DHchiral} indicate that enslavement,  $\bs= s\hbp$ in (\ref{enslavement}), is  \emph{not preserved}~:  
 \emph{spin can not be consistently enslaved} within the minimal  S-model which is therefore definitely
\emph{different} from the c-model.
This is also obvious by counting the degrees of freedom: the electromagnetic field breaks the Wigner-Souriau translations  and
  reduces the dimension of the kernel from $3$ to $1$, therefore the space of motions is $8$, and not  $6$-dimensional. Unlike as in the free case,
\emph{spin is a genuine degree of freedom, which can not be eliminated}. 

In the non-singular case yet another tedious calculation allows 
us to find the  Poisson brackets,
\beq\hskip1mm
\barraynb{lllll}
\{r_i,r_j \}&=&-\displaystyle\frac{\epsilon_{ijk}\hat{p}_k}{e\hat{p}\cdot \bB},  
&\{r_i, p_j\}=\displaystyle\frac{B_i\hat{p}_j}{\hat{p}\cdot \bB}, 
&\hskip-15mm\{p_i,p_j\}= 0, 
\\[12pt]
\{s_i,r_j \}&=&\displaystyle\frac{|\bp|}{e\hat{\bp}\cdot\bB}(-\delta_{ij}+\hat{p}_i\hat{p}_j),\qquad
&\{s_i,p_j\}=\displaystyle\frac{|\bp|}{\hat{\bp}\cdot\bB}\big(\epsilon_{ijk}B_k+\hat{p}_i(\hat{\bp}\times\bB)_j\big),
\\[12pt]
\{s_1,s_2\}&=&s_3-\displaystyle\frac{|\bp|p_3}{e\hat{\bp}\cdot\bB}
&&
\earraynb
\label{SPB}
\eeq
which are  substantially different from those for the c-model, (\ref{cPB}). 
The Jacobi identities follow from $d\omega_S=0$, and can also be checked directly.
Note here the absence of the usual ``Heisenberg''  term $\delta_{ij}$ in  $\{r_i,p_j\}$, similar to 
the dropping out of the momentum term $\hbp$  from the velocity relation in (\ref{Seqsa}). Note also that the momenta commute instead of closing on the magnetic field, as expected.  
It is therefore reassuring that
 the associated Hamilton equations yield (\ref{Seqsa})-(\ref{Seqsb})-(\ref{Seqsb}) as they should. 
Thus, the difference between the coupled c- and S-models originates in  both the symplectic structure and the Hamiltonian.
\goodbreak

\section{Chiral Magnetic Effect and Chiral Anomaly}

The particle current is determined in terms of the coordinates $\br, \bp, s_1, s_2$ using the determinant of the symplectic matrix (\ref{Sdet}).   
The invariant phase space volume element of the 8-dimensional space of motions is ${\rm V}_S=\omega_S^4/4!=\sigma_S^4/4!$ 
\cite{SSD,DHHMS} i.e., by (\ref{reducedomega}),
\beq
\label{Smeasure}
d{\rm V}_S=D_S\,d^3{\br}d^3{\bp}ds_1ds_2=
\frac{eB}{|\bp|^2}\,d^3{\br}\,d^3{\bp}\,ds_1ds_2
\eeq
also expressed in a covariant way useful for calculations, 
\beq
\label{cSmeasure}
d{\rm V}_S=
\frac{e\hat{\bp}\cdot\bB}{|\bp|^2}\,\delta\big(\hat{\bp}\cdot\bs-\frac{1}{2}\big)\,d^3{\br}d^3{\bp}d^3\bs\,.
\,
\eeq

If $f(\br,\bp, s_1, s_2 )$ is a distribution function on the spin-extended space of motions, which we assume to satisfy again the collisionless Boltzmann equation, now
$
\displaystyle\frac{\partial f}{\partial t}+\dot{r}_i\displaystyle\frac{\partial f}{\partial r_i}+\dot{p}_i\displaystyle\frac{\partial f}{\partial p_i}+\dot{s}_a\frac{\partial f}{\partial s_a}=0 
$ ($i=1,2,3,\ a=1,2$), the
 the particle current,
\begin{eqnarray}
\label{Scurrent}
\bm{j}(\br, t)\!&=&\!\!\int\!\! f\dot{\br}\,D_Sd^3\bp\, ds_1ds_2
=
e\bB\int\! \frac{f}{|\bp|^2\hat{p}_3}d^3\bp\, ds_1ds_2
+e{\bE}\times\int\! \frac{f\hat{\bp}}{|\bp|^2\hat{p}_3}d^3\bp\, ds_1ds_2\qquad
\end{eqnarray}
is decomposed  into  merely \emph{two} (and not three) terms, namely into a chiral- magnetic and an anomalous current analogous to those, (\ref{ccurrent}), for the c-model \cite{StephanovYin,SonYama2}. The absence of a normal current follows from that of the usual $\hbp$. 
The particle density 
$
\rho(\br, t)=\displaystyle\int{f}\, D_Sd^3\bp\, ds_1 ds_2 ,
$
satisfies 
\beq
\frac{\partial \rho(\br, t)}{\partial t}=\int \Big(\frac{\partial{D_S}}{\partial t}
 \Big)fd^3\bp ds_1 ds_2
+\int D_S
 \frac{\partial f}{\partial t} d^3\bp ds_1 ds_2.
\eeq 
Dropping boundary terms we find, for constant fields $\bB, \bE$ and no explicit time dependence,   
\beqa
\frac{\partial \rho(\br,t)}{\partial t}+\bm{\nabla}_{\br}\cdot\bm{j}(\br,t)&=&e^2\bE\cdot\bB
\int\! f\bm{\nabla}_\bp\cdot \big(\frac{\hat{\bp}}{|\bp|p_3}\big)d^3\bp ds_1ds_2=
e^2\bE\cdot\bB\, \frac{4\pi\,f_0}{\hat{p}_3},\quad
\label{Sanomaly}
\eeqa
which differs from (\ref{canomaly}) valid for the c-model  by the same 
 factors as (\ref{SLiouville}) does from (\ref{cLiouville}) and $f_0=\displaystyle\int\! f(\br,\bp=0, s_1, s_2)\, ds_1 ds_2$.

\section{Conclusion}

In this Letter we demonstrated that
 the spin-extended chiral model (our S-model), instrumental in deriving the twisted Lorentz symmetry of the c-model  is equivalent to the latter only in the free case but not when coupling to an external field is considered.
 This is highlighted by the inconsistency with the S-dynamics with  ``enslaving''.
 
The difference comes from the  different choice  of what one considers as ``position'': the  c-model is coupled to the e.m. field by viewing $\bx$ in (\ref{comega}) as a position \cite{StephanovYin,SonYama2,Stone,QunWang,Manuel}, with no attention paid at its ``twisted'' behavior under a Lorentz boost \cite{ChenSon,DHchiral,deAzcarraga,Kosinski}. 
  In the  S-model instead, the coupling is introduced in terms of the ``true'' position, $\br$, which does transform in the usual way under a Lorentz boost \cite{DHchiral}.  
 Let us emphasize that the S-model and its coupling to an external field follow from First Principles -- namely of Souriau's Mechanics \cite{SSD}. We stress that our ``true position'', $\br$, is defined on the evolution space $V^9$ and \emph{not} on $M$: it is the \emph{combination} of position, time, momentum and spin,
\beq
\tilde{\bx}=\frac{\bg}{|\bp|}=
 \br-\hbp\,{}t+\frac{\hbp\times\bs}{|\bp|}
\label{tbx} 
\eeq
which is conserved and can label a point of $M$ -- i.e., a motion. The  Poincar\'e group acts naturally on $V^9$, whereas its action projected onto the space of motions  is ``twisted'' and is \emph{not} natural \cite{DHchiral}. 

We note also the the expression (\ref{Sangmom}) of the angular momentum ``lives" on the evolution space. Using the space of motion coordinates  $\tbx$ and $\bp$ allows us to absorb the evolution space coordinates into conserved quantities and convert the angular momentum it into
\beq
\belle=
 \tbx\times\bp\,+\half\hbp.
 \label{xSangmom}
\eeq 
In particular the ``unchained'' part of the spin vector $\bs$ has been absorbed into $\tbx$, leaving us with the ``enslaved'' contribution $\half\hbp$.

The  $\bx$ used in the c-model is in turn a  \emph{label of c-motions} obtained by putting $t=0$ into  $\bx(t)=\bx+\hbp{t}$, obtained by integrating the  c-equations of motion, (\ref{ceqs}), in the free case
with initial condition $\bx(t)=\bx$. Therefore  viewing $\bx$ in the c-model as a ``position,'' is, in our opinion, unjustified. 

Returning to the coupled case,  we mention that
it has been suggested  \cite{SonYama2,ChenSon,Manuel} to modify the c-Hamiltonian by adding a term, 
\beq
h \to |\bp| -e\frac{\hbp\cdot\bB}{2|\bp|}.
\label{ChenSonHam}
\eeq
Such a modification is certainly possible and can  be  generalized to the spin-dependent case \cite{DHchiral}. Anomalous
coupling yields in fact the dispersion relation
\beq
\cE = \sqrt{|\bp|^2 -(eg/2)S\cdot F},
\qquad
S\cdot F= \bs\left(\bB-\frac{\bp}{\cE}\times \bE\right).
\label{DHdispert}
\eeq
where the real number $g$ represents the gyromagnetic ratio  \cite{DHchiral}.
For $s=\half$ $g=2$ and a weak purely magnetic field,  (\ref{DHdispert})
approximately reduces  to (\ref{ChenSonHam}). In this Letter, we studied the minimal case $g=0$. Our study will be extended to anomalous coupling
elsewhere.

The S-model exhibits properties which are similar to those the chiral one, namely the chiral anomaly, CME and AHE. Its advantage is its manifest Lorentz invariance.

Remarkably, the anomaly vanishes precisely when the system carries a full Poincar\'e symmetry -- namely when the Pfaffian invariant vanishes.


We just mention  that the additional spin degree of freedom would allow us also consider the \emph{spin current} defined, by analogy to 
(\ref{Scurrent}), as
\beqa
\bj_s\! &=&\!\displaystyle\int\!{f\dot{\bs}}\, D_Sd^3\bp\, ds_1ds_2
\nn\\[4pt]
&=&
-e\int\!\frac{f\bp}{|\bp|^2\hat{p}_3}(\hbp\cdot\bE)\,{d^3\bp\, ds_1ds_2}
-
e\bB\times\int\!\frac{f\bp}{|\bp|^2\hat{p}_3}\,{d^3\bp ds_1ds_2}
+
e\bE\int\!\frac{f|\bp|}{|\bp|^2\hat{p}_3}\,{d^3\bp\, ds_1ds_2}.\qquad\quad
\label{spincurrent}
\eeqa
We have here again three terms as in (\ref{ccurrent}). Note also the the magnetic and anomalous Hall currents are sort of duals to those in
(\ref{Scurrent}) in that $\bB \to \bE$ and $\bE \to -\bB$. The  spin current and its relation to QED is under current study.

\begin{acknowledgments} 
Special thanks are due to Christian Duval, who allowed us to reproduce his remarks \cite{CDPriv} on full Poincar\'e symmetry,  and to Pengming Zhang for helping us to produce our figure.
We would also like to thank \"Omer Faruk Dayi, Gary Gibbons, Mike Stone and Juan Torres-Rincon for discussions correspondence. ME is also indebted to Berkin Malkoc for his support and Miletos Inc. for hospitality, where part of this work was completed. 
This research was supported by the Major State Basic Research Development Program in China (No.
2015CB856903) and the National Natural Science Foundation of
China (Grant No. 11035006 and 11175215). We are both grateful to the Institute of Modern Physics of the Chinese Academy of Sciences in Lanzhou  for hospitality extended to us.
\end{acknowledgments}
\goodbreak



\end{document}